\documentclass[a4paper,10pt]{llncs}
\usepackage{graphicx}  
\usepackage{subfigure}

\title{Adaptive Thresholds for Layered Neural Networks with Synaptic Noise}
\author{D. Boll\'e and R. Heylen}
\institute{Institute for Theoretical Physics\\
Katholieke Universiteit Leuven, Celestijnenlaan 200 D\\
B-3001, Leuven, Belgium}

\begin{document}

\maketitle

\begin{abstract}
The inclusion of a macroscopic adaptive threshold is studied for the retrieval dynamics of layered feedforward neural network models with synaptic noise. It is shown that if the threshold is chosen appropriately as a function of the cross-talk noise and of the activity 
of the stored patterns, adapting itself automatically in the course of
the recall process, an autonomous functioning of the network is guaranteed.
This self-control mechanism considerably improves the quality of retrieval, in particular 
the storage capacity, the basins of attraction and the mutual information content.

\end{abstract}

\section{Introduction}

As is common knowledge by now, layered feedforward neural network models are the workhorses in many practical applications in several areas of research and, therefore, any new insight in their capabilities and limitations should thus be welcome.
In view of the fact that in many of these applications, e.g., pattern recognition in general, information is mostly encoded by a small fraction of bits and that also in neurophysiological studies the activity level of real neurons is found to be low, any reasonable network model has to allow variable activity of the neurons. The limit of low activity, i.e., sparse coding is then especially interesting. 
Indeed, sparsely coded models have a very large storage capacity behaving as
$1/(a\ln a)$ for small $a$, where $a$ is the 
activity (see, e.g., \cite{W,P,Ga,Ok} and references therein). However,
for low activity the basins of attraction might become very small and
the information content in a single pattern is reduced \cite{Ok}.
Therefore, the necessity of a control of the activity of the neurons has
been emphasized such that the latter stays the same as the activity of the
stored patterns during the recall process.
This has led to several discussions imposing external constraints on the
dynamics of the network. However, the enforcement of such a constraint at every time
step destroys part of the autonomous functioning of the network, i.e., a functioning that has to be independent precisely from such external constraints or control mechanisms.
To solve this problem, quite recently  a self-control mechanism has been
introduced in the dynamics of networks for so-called diluted architectures \cite{DB98}. This self-control mechanism introduces  a time-dependent 
threshold in the transfer function \cite{DB98,BDA00}. It is determined as a 
function of both the cross-talk noise and the
activity of the stored patterns in the network, and adapts itself in
the course of the recall process. It furthermore allows to reach optimal retrieval performance both in the absence and in the presence of synaptic noise \cite{DB98,BDA00,BH04,DKTE02}. These diluted architectures contain no common ancestors nodes, in contrast with feedforward architectures. It has then been shown that a similar mechanism can be introduced succesfully for layered feedforward architectures but, without synaptic noise~\cite{BM00}.

The purpose of the present contribution is 
to generalise this self-control mechanism for layered architectures when synaptic noise is allowed, and to show that it leads to a substantial improvement of the quality of 
retrieval, in particular the storage capacity, 
the basins of attraction and the mutual information content.

\section{The model}

Consider a neural network composed of binary neurons arranged in layers, each
layer containing $N$ neurons. A neuron can take values ~$\sigma_{i}(t) \in
\{0,1\}$ where $t=1,\ldots,L$ is the layer index and~ $i=1, \ldots ,N$~
labels the neurons. Each neuron on layer $t$ is unidirectionally
connected to all neurons on layer $t+1$.
We want to memorize $p$ patterns  $\{\xi_i^\mu(t)\},
{{i=1,\ldots,N}, ~{\mu=1,\ldots,p}}$ on each layer $t$, taking the
values $\{0,1\}$. They are assumed to be
independent identically distributed random variables (i.i.d.r.v.) with respect
to $i$, $\mu$ and $t$, determined by the probability distribution:
$p(\xi_i^\mu (t))=a\delta(\xi_i^\mu (t)-1)+(1-a)\delta(\xi_i^\mu (t))$.
From this form we find that the
expectation value and the variance of the patterns are given by 
   $
     E[\xi_i^\mu (t)]=E[\xi_i^\mu (t)^2]=a~.
   $
Moreover, no statistical correlations occur, in fact for $\mu\neq\nu$
the covariance vanishes. 

The state $\sigma_{i}(t+1)$ of neuron $i$ on layer $t+1$ is determined
by the state of the neurons on the previous layer $t$ according to the  
stochastic rule 
\begin{equation}  
       \label{eq:stoc}
  P(\sigma_{i}(t+1)\mid \sigma_{1}(t), \ldots ,\sigma_{N}(t))
             = \{1+\exp[2(2\sigma_i(t+1)-1) \beta{h_i(t)}]\}^{-1}.
\end{equation}
The right hand side is the logistic function. The ``temperature" $T=1/\beta$ controls the stochasticity of the network dynamics, it measures the synaptic noise level~\cite{HKP91}.
Given the network state  $\{\sigma_i(t)\};{i=1,\ldots,N}$ on layer $t$,
the so-called ``local field" ${h_i(t)}$ of neuron $i$ on the next layer $t+1$ is given by 
\begin{equation} 
       \label{eq:h}
   h_i(t)= \sum_{j=1}^{N} 
          J_{ij}(t)(\sigma_j(t) -a)-\theta(t) 
\end{equation} 
with $\theta(t)$ the threshold to be specified later. 
The couplings $J_{ij}(t)$ are the synaptic strengths of the interaction  
between neuron $j$ on layer $t$ and neuron $i$ on layer $t+1$. They
depend on the stored patterns at different layers according to the
covariance rule 
\begin{equation}   
       \label{eq:j}
   J_{ij}(t)=\frac{1}{N {a(1-a)}} \sum_{\mu=1}^{N} 
             (\xi_i^\mu (t+1)-a)(\xi_j^\mu (t)-a)~.
\end{equation}
These couplings then permit to store sets of patterns to be retrieved by
the layered network.

The dynamics of this network is defined as follows (see \cite{DKM}).
Initially the first layer (the input) is externally set in some fixed
state. In response to that, all neurons of the second layer update
synchronously at the next time step, according to the stochastic rule
(\ref{eq:stoc}), and so on.

At this point we remark that the couplings (\ref{eq:j}) are of
infinite range (each neuron interacts with infinitely many others) such
that our model allows a so-called mean-field theory approximation. This 
essentially means that we focus on the dynamics of a single neuron
while replacing all the other neurons by an average background local
field. In other words, no fluctuations of the other neurons are taken into
account.
In our case this approximation becomes exact because, crudely speaking, 
$h_{i}(t)$ is the sum of very many terms and a central limit theorem can
be applied \cite{HKP91}. 

It is standard knowledge by now that mean-field theory
dynamics can be solved exactly for these layered architectures 
(e.g., \cite{DKM,B04}). 
By exact analytic treatment we mean that, given the state of the
first layer as initial state, the state on layer $t$ that
results from the dynamics is predicted by recursion formulas.
This is essentially due to the fact that the     
representations of the patterns on different layers are chosen independently. 
Hence, the big advantage is that this 
will allow us to determine the  effects from self-control in an 
exact way. 

The relevant parameters describing the solution
of this dynamics are the 
{\it main overlap} of the  state of the network and the $\mu$-th
pattern, and the {\it neural activity} of the neurons   
\begin{equation}
    \label{M(t)}
  M^\mu(t)  =
    \frac{1}{N{a(1-a)}}\sum_{i=1}^N{(\xi_i^\mu(t)-a)}
                     (\sigma_i(t) -a),
    \qquad
   q(t) =
      \frac{1}{N}\sum_{i=1}^N \sigma_i(t)~.
\end{equation} 

In order to measure the retrieval quality of the recall process, we use
the mutual information function \cite{DB98,BDA00,NBP98,ST98}. In general, it
measures the average amount of information that can be received by
the user by observing the signal at the output of a channel \cite{B90,S48}.
For the recall process of stored patterns that we are discussing 
here, at each layer the process can be regarded as a channel with
input $\xi_i^\mu(t)$ and output $\sigma_{i}(t)$ such that this mutual
information function can be defined as \cite{DB98,B90}
\begin{equation}  
    \label{eq:inf}
   I(\sigma_i(t);\xi_i^\mu (t))=
       S(\sigma_i(t))-\langle S(\sigma_i(t)|\xi_i^\mu (t))\rangle
                                 _{\xi^{\mu}(t)}
\end{equation}
where  $S(\sigma_i(t))$ and $S(\sigma_i(t)|\xi_i^\mu (t))$ are the entropy
and the conditional entropy of the output, respectively
\begin{eqnarray} 
      \label{eq:en}
  S(\sigma_i(t))&=& -\sum_{\sigma_i} p(\sigma_i(t))\ln[p(\sigma_i(t))]\\ 
      \label{eq:enc}
  S(\sigma_i(t)|\xi_i^\mu (t))&=&
       -\sum_{\sigma_i} p(\sigma_i(t)|\xi_i^\mu (t))
                \ln[p(\sigma_i(t)|\xi_i^\mu (t))]~.
\end{eqnarray}
These information entropies are peculiar to the probability 
distributions of the output. 
The quantity $p(\sigma_i(t))$ denotes the probability distribution for the
neurons at layer $t$ and $p(\sigma_i(t)|\xi_i^\mu (t))$ indicates the
conditional probability that the $i$-th neuron is in a state
$\sigma_i(t)$ at layer $t$ given that the $i$-th site of the 
pattern to be  retrieved is $\xi_i^\mu (t)$.
Hereby, we have assumed that the conditional probability of all the
neurons factorizes, i.e.,
 $p(\{\sigma_i(t)\}|\{\xi_i(t)\})=\prod_j p(\sigma_j(t)|\xi_j(t))$, which is a
consequence of the mean-field theory character of our model explained
above. We remark that a similar factorization  has also been
used in Schwenker et al.~\cite{SSP96}.

The calculation of the different terms in the expression (\ref{eq:inf})
proceeds as follows. Because of the mean-field character of our model the following formula hold for every neuron $i$ on each layer $t$. Formally writing (forgetting about the pattern index $\mu$) $\langle O \rangle
\equiv \langle \langle O \rangle_{\sigma|\xi} \rangle_{\xi}=
\sum_{\xi} p(\xi) \sum_{\sigma} p(\sigma|\xi) O $ for an arbitrary
quantity $O$ the conditional probability can be obtained in a rather
straightforward way by using the complete knowledge about the system:
$\langle \xi \rangle=a, \, \langle \sigma \rangle=q, \,
\langle (\sigma-a)( \xi-a) \rangle=M, \, \langle 1 \rangle=1$.

The result reads
\begin{equation}
    p(\sigma|\xi)=[\gamma_0\xi+(\gamma_1-\gamma_0)\xi]\delta(\sigma-1)+
       [1-\gamma_0-(\gamma_1-\gamma_0)\xi]\delta(\sigma)
\end{equation}
where $\gamma_0=q-aM$ and $\gamma_1=(1-a)M+q$, and where the $M$
and $q$ are precisely the relevant parameters (\ref{M(t)}) for large $N$.
Using the probability distribution of the patterns we obtain
\begin{equation}
   p(\sigma)=q\delta(\sigma-1)+(1-q)\delta(\sigma)~.
\end{equation}   
Hence the entropy (\ref{eq:en}) and the conditional entropy
(\ref{eq:enc})  become
\begin{eqnarray}
           S(\sigma)=&-&q\ln q -(1-q)\ln(1-q) \\
           S(\sigma|\xi)=&-&[\gamma_0+(\gamma_1-\gamma_0)\xi]
	             \ln[\gamma_0+(\gamma_1-\gamma_0)\xi] 
		     \nonumber \\
                &-&[1-\gamma_0-(\gamma_1-\gamma_0)\xi]
                      \ln[1-\gamma_0-(\gamma_1-\gamma_0)\xi]~.
\end{eqnarray}		      
By averaging the conditional entropy over the pattern $\xi$ we finally get
for the mutual information function (\ref{eq:inf}) for the layered model
\begin{eqnarray}
     \label{eq:Ifin}
     I(\sigma;\xi) = -q\ln q -(1-q)\ln(1-q)
          +a[\gamma_1\ln\gamma_1+(1-\gamma_1)\ln(1-\gamma_1)]
	  \nonumber\\
	     +(1-a)[\gamma_0\ln\gamma_0+(1-\gamma_0)\ln(1-\gamma_0)]~. 
\end{eqnarray}

\section{Adaptive thresholds}

It is standard knowledge (e.g., \cite{DKM})
that the synchronous dynamics for layered  architectures 
can be solved exactly following the method based upon a signal-to-noise
analysis of the local field (\ref{eq:h}) (e.g., \cite{Ok,B04,A77,AM88} and references therein). 
Without loss of generality we focus on the recall of one pattern, say
$\mu=1$, meaning that only $M^1(t)$ is macroscopic, i.e., of order $1$
and the rest of the patterns causes a cross-talk noise at each 
step of the dynamics.

We suppose that the initial state of the network model $\{\sigma_i(1)\}$ 
is a collection of i.i.d.r.v. with average and variance given by 
$
      E[\sigma_i(1)]=E[(\sigma_i(1))^2]=q_0~.
$
We furthermore assume that this state is correlated with only
one stored pattern, say pattern $\mu=1$, such that
 $  \mbox{\rm Cov}(\xi_i^\mu (1),\sigma_i(1))=\delta_{\mu,1}~M_0^1~{a(1-a)}~.$

Then the full recall proces is described by \cite{DKM,B04}
\begin{eqnarray} 
&&   M^1(t+1) =  \frac{1}{2}\left\{\int{\cal D} x
       \tanh\left[\beta((1-a)M^1(t)-\theta(t)+\sqrt{\alpha D(t)}\,x)\right]
             \right. \nonumber \\  
	  && \hspace*{1cm}  +  \left.  \int{\cal D} x
       \tanh\left[\beta(-aM^1(t)-\theta(t)+ \sqrt{\alpha D(t)}\,x)\right]
                   \right\} 
              \label{eq:a} \\
 &&    q(t+1)= aM^1(t+1) \nonumber \\
&&       \hspace*{1cm} +\frac{1}{2} 
           \left\{ 1+\int{\cal D} x
     \tanh\left[\beta(-aM^1(t)-\theta(t)+\sqrt{\alpha D(t)}\,x)\right]
                        \right\}
         \label{eq:b}  \\
 &&    D(t+1)= Q(t+1) \nonumber\\
 &&       \hspace*{1cm}  + 
        \frac{\beta}{2}\left\{1-a\int{\cal D} x\tanh^2\beta\left[(1-a)M^1(t)
            -\theta(t)+\sqrt{\alpha D(t)}\,x\right] \right. 
	    \nonumber \\
	 && \hspace*{1cm} - \left. (1-a)\int{\cal D} x\tanh^2\beta
	\left[-aM^1(t)-\theta(t)+\sqrt{\alpha D(t)}\,x\right]\right\}^2 D(t)
	\label{eq:c}
\end{eqnarray}  
where $\alpha= p/N$, ${\cal D} x$ is the Gaussian measure ${\cal D} x= dx
(2\pi)^{-1/2}\exp(-x^2/2)$, where $Q(t)=[(1-2a)q(t)+a^2]$ and where $D(t)$ contains the influence of the cross-talk noise caused by the patterns $\mu>1$. 
As mentioned before,  $\theta(t)$ is an adaptive threshold that has to be chosen. 

In the sequel we discuss two different choices and both will be compared for
networks with synaptic noise and various activities.  Of course, it is known that the quality of the recall process is influenced by the cross-talk noise. An idea is then to introduce a threshold that adapts itself autonomously in the course of the recall process and that counters, at each layer, the cross-talk noise. This is the self-control method proposed in \cite{DB98}. This has been studied for layered neural network models  without
synaptic noise, i.e., at $T=0$, where the rule (\ref{eq:stoc}) reduces  to the deterministic form 
  $  \sigma_i(t+1)=\Theta({h_i(t)}) $
with $\Theta(x)$ the Heaviside function taking the value $\{0,1\}$.
For sparsely coded models, meaning that the pattern activity $a$ is very
small and tends to zero for $N$ large, it has been found \cite{BM00} that 
\begin{equation}   
    \label{eq:thr}
     \theta(t)_{sc}= c(a)\sqrt{\alpha D(t)}, \quad  c(a)=\sqrt{-2\ln a}
\end{equation}
makes the second term on the r.h.s of Eq.(\ref{eq:b}) at $T=0$, asymptotically vanish
faster than $a$ such that $q \sim a$. 
It turns out that the inclusion of this self-control threshold  considerably improves the quality of retrieval, in particular the storage capacity, the basins of attraction and the information content.

\begin{figure}[htp]
\centering
\includegraphics[width=6.5cm]{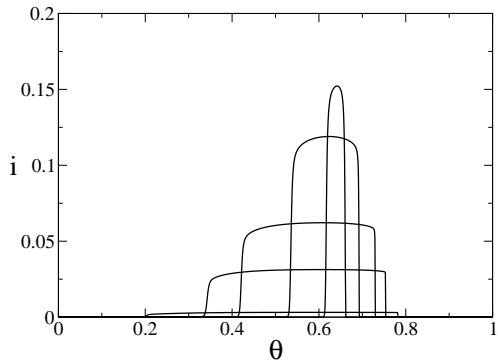}
\caption{ The information $i=\alpha I$ as a function of $\theta$ for $a=0.005$, $T=0.1$ and several values of the load parameter $\alpha=0.1,1,2,4,6$ (bottom to top)}
\label{fig:infoth}
\end{figure}

The second approach chooses a threshold by maximizing the information
content, $i=\alpha I$ of the network (recall Eq.~(\ref{eq:Ifin})). This function  depends on $M^1(t)$, $q(t)$,  $a$, $\alpha$ and $\beta$. The
evolution of $M^1(t)$ and of $q(t)$ (\ref{eq:a}), (\ref{eq:b}) depends on the specific choice of the threshold through the local field (\ref{eq:h}). We consider a
layer independent threshold $\theta(t)=\theta$  and calculate the value of 
(\ref{eq:Ifin}) for fixed $a$, $\alpha$, $M_0^1$, $q_0$ and $\beta$. The optimal 
threshold, $\theta=\theta_{opt}$, is then
the one for which the mutual information function is maximal.
The latter is non-trivial
because it is even rather difficult, especially in the limit of sparse
coding, to choose a threshold interval by hand such that $i$ is 
non-zero. The computational cost will thus be larger compared to the one of the self-control approach. To illustrate this we plot in Figure~\ref{fig:infoth} the information content $i$ 
as a function of $\theta$  without self-control or a priori optimization, for $a=0.005$ and different values of $\alpha$.
For every value of $\alpha$, below its critical value, there is a range for the threshold 
where the information content is different from zero and hence, retrieval is possible. This
retrieval range becomes very small when the storage capacity approaches its
critical value $\alpha_c=6.4$. 

Concerning then the self-control approach, the next problem to be posed in analogy with the case without synaptic noise is the following one. Can one determine a form for
the threshold $\theta(t)$ such that the integral in the second term on the r.h.s of Eq.(\ref{eq:b}) at $T \neq 0$ vanishes asymptotically faster than $a$? 

In contrast with the case at zero temperature where due to the simple form of the transfer function, this threshold could be determined analytically
(recall Eq.~(\ref{eq:thr})), a  detailed study of the asymptotics of the 
integral in Eq.~(\ref{eq:b}) gives no satisfactory analytic solution. 
Therefore, we have designed a systematic numerical procedure through the
following steps: 
\begin{itemize}
\item Choose a small value for the activity $a'$.
\item Determine through numerical integration the threshold $\theta'$
such that 
\begin{equation}
\int_{-\infty}^{\infty} 
     \frac{dx \,\,e^{-x^2/ 2 \sigma^2}}{\sigma \sqrt{2\pi }} 
 \Theta (x- \theta) \leq a' \quad \mbox{for} \quad \theta > \theta' 
\end{equation}
for different values of the variance $\sigma^2={\alpha D(t)}$.
\item Determine as a function of  $T=1/\beta$, the value
for $\theta'_T$ such that  
\begin{equation}
 \int_{-\infty}^{\infty} 
     \frac{dx \,\,e^{-y^2/ \sigma^2}}{2 \sigma \sqrt{2\pi }} 
  [1+ \tanh[\beta (x- \theta)]] \leq a'  \quad
                 \mbox{for} \quad \theta > \theta' +\theta'_T.
\end{equation}
\end{itemize}
The second step leads precisely to a threshold having the form of Eq.~(\ref{eq:thr}). The 
third step determining the temperature-dependent part $\theta'_T$ leads to the final 
proposal
\begin{equation}
    \theta_{t}(a,T)=\sqrt{-2 \ln (a)\alpha D(t)} - \frac12 \ln(a) T^2.
     \label{threstemp}
\end{equation}
This dynamical threshold is again a
macroscopic parameter, thus no average must be taken over the microscopic
random variables at each step $t$ of the recall process.

\begin{figure}[htp]
\centering
\includegraphics[width=6cm]{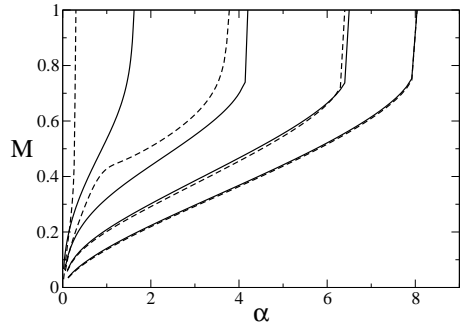}
\caption{The basin of attraction as a function of $\alpha$ for $a=0.005$
and $T=0.2, 0.15, 0.1, 0.05$ (from left to right) with 
(full lines) and without (dashed lines) the $T$-dependent part 
in the threshold (\ref{threstemp}).}
\label{fig:basinsT}
\end{figure}

\begin{figure}[htp]
\centering
\includegraphics[height=4.2cm,width=10cm]{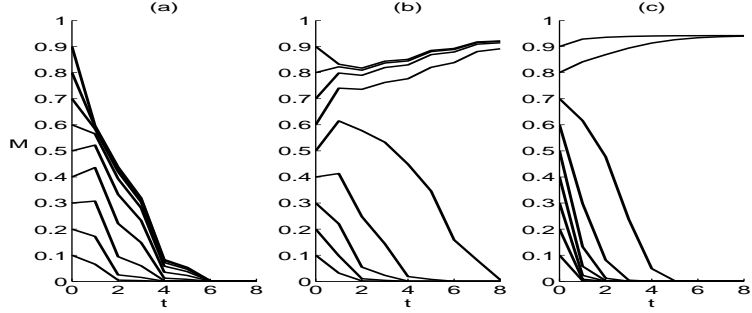}
\caption{The evolution of the main overlap $M(t)$ for
	     several initial values $M_0$ with $T=0.2,~q_0=a=0.005,~\alpha=1$
	     for the self-control model (\ref{threstemp}) without (a) and with $T$-dependent part (b) and for the optimal threshold model (c).}
\label{fig:evolM}
\end{figure}

\begin{figure}[htp]
\centering
\includegraphics[width=6.5cm]{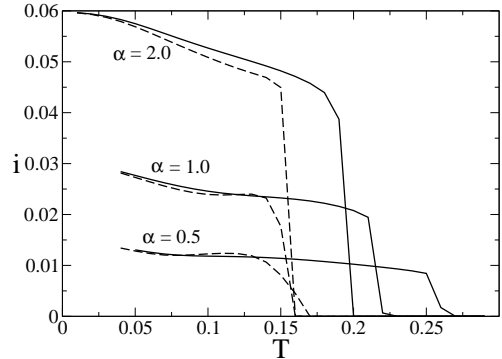}
\caption{The information content $i=\alpha I$ as a function of $T$ for several
values of the loading $\alpha$ and $a=0.005$ with (full
lines) and without (dashed lines) the $T$-correction 
in the threshold.}
\label{fig:infoT}
\end{figure}

We have solved these self-controlled dynamics,
Eqs.(\ref{eq:a})-(\ref{eq:c}) and (\ref{threstemp}), for our model with
synaptic noise, in the limit of sparse coding, numerically. In
particular, we have studied in detail the influence
of the $T$-dependent part of the threshold. Of course,
we are only interested in the retrieval solutions with $M>0$ (we forget about the index $1$) and
carrying a non-zero information~$i=\alpha I$. 
The important features of the solution are illustrated, for a typical value of $a$ in Figures~\ref{fig:basinsT}-\ref{fig:infoT}. 
In Figure~\ref{fig:basinsT} we show
the basin of attraction for the whole retrieval phase for the model with
threshold (\ref{eq:thr}) (dashed curves) compared to 
the model with the noise-dependent threshold (\ref{threstemp})
(full curves). We see that there is no clear improvement for low $T$
but there is a substantial one for higher $T$.
Even near the border of critical storage the results are still improved
such that also the storage capacity itself is larger. 

This is further illustrated in Figure~\ref{fig:evolM} where we compare
the evolution of the retrieval overlap $M(t)$ starting from several initial
values, $M_0$, for the model with (Figure \ref{fig:evolM} (a)) and without 
(Figure \ref{fig:evolM} (b)) the $T$-correction in the threshold and for the optimal threshold model (Figure \ref{fig:evolM} (c)). Here this 
temperature correction is absolutely crucial to guarantee retrieval, i.e., $M \approx 1$. It really makes the difference between retrieval and non-retrieval in the model. 
Furthermore, the model with the self-control threshold with noise-correction has even a wider basin of attraction than the model with optimal threshold.

In Figure~\ref{fig:infoT} we plot the information content $i$ as a
function of the temperature for the self-control dynamics with the 
threshold (\ref{threstemp}) (full curves), respectively (\ref{eq:thr})
(dashed curves). We see that a substantial improvement of the
information content is obtained.  

\begin{figure}
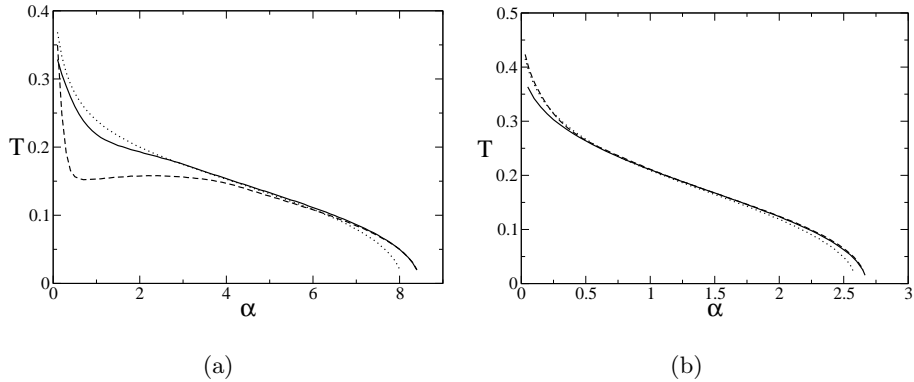

       \centering   
       \subfigure[]{
             \label{fig:phases1}
                 \includegraphics[width=5.8cm]{phases_a0.005.eps}}
	 \hspace{0.1cm}
	\subfigure[]{
	\label{fig:phases2}
             \includegraphics[width=5.8cm]{phases_a0.02.eps}}
\caption{Phases in the $T-\alpha$ plane for $a=0.005$ (a) and $a=0.02$ (b) with (full
line) and without (dashed line) the temperature correction 
in the self-control threshold and with optimal threshold (dotted line).}
  \label{fig:phases}
\end{figure}

Finally we show in Figure~\ref{fig:phases} a $T-\alpha$ plot for $a=0.005$ (a) and $a=0.02$ (b) with (full line) and without (dashed line) noise-correction 
in the self-control threshold and with optimal threshold (dotted line).
These lines indicate two phases of the layered model: below the lines our model allows recall, above the lines it does not. For $a=0.005$ we see that the $T$-dependent term in the self-control threshold leads to a big improvement in the region for large noise and small loading and in the region of critical loading. For $a=0.02$ the results for the self-control threshold with and without noise-correction and those for the optimal thresholds almost coincide, but we recall that the calculation with self-control is autonomously done by the network and less demanding computationally.

\section{Conclusions}
In this work we have studied the inclusion of an adaptive threshold in 
sparsely coded layered neural networks with synaptic noise. 
We have presented an analytic form for a self-control threshold, allowing an autonomous functioning of the network, and compared it with an optimal threshold obtained by maximizing the mutual information which has to be calculated externally each time one of the network parameters (activity, loading, temperature) is changed.
The consequences of this self-control mechanism on the quality of the recall process have been studied.

We find that the basins of attraction of the retrieval solutions as well
as the storage capacity are enlarged. For some activities the self-control threshold even sets the border between retrieval and non-retrieval.
This confirms the considerable improvement of the quality of recall by 
self-control, also for layered network models with synaptic noise. 

This allows us to conjecture that 
self-control might be relevant for other architectures in the presence
of synaptic noise, and even for dynamical systems in general, when
trying to improve, e.g.,  basins of attraction .\\

\noindent
{\bf Acknowledgment}\\

\noindent 
This work has been supported by the Fund for Scientific Research- Flanders
(Belgium).


\begin{thebibliography}{1}

\bibitem{W} Willshaw D J, Buneman O P, and Longuet-Higgins H C,
         Nonholographic associative memory, 
      {\it Nature} {\bf 222} (1969) 960.  
\bibitem{P} Palm G,  
   On the storage capacity of an associative memory with random
       distributed storage elements, 
      {\it Biol. Cyber.} {\bf 39} (1981) 125. 
\bibitem{Ga} Gardner E, 
    The space of interactions in neural network models,
       {\it J. Phys. A: Math. Gen.} {\bf 21} (1988) 257. 
\bibitem{Ok} Okada M, 
       Notions of associative memory and sparse coding, 
        {\it Neural Networks} {\bf 9} (1996) 1429.
\bibitem{DB98} Dominguez D R C and Boll\'e D,  
       Self-control in sparsely  coded networks, 
     {\it Phys. Rev. Lett.} {\bf 80} (1998) 2961.  
\bibitem{BDA00} Boll\'e D, Dominguez D R C  and  Amari S,  
       Mutual information of sparsely coded associative memory with self-control and
       ternary neurons,
        {\it Neural Networks} {\bf 13}(2000) 455.
\bibitem{BH04} Boll\'e D  and Heylen R,
       Self-control dynamics for sparsely coded networks with synaptic noise, in
      {\it 2004 Proceedings of the IEEE International Joint Conference on Neural Networks}, 
      p.3195
\bibitem{DKTE02}  Dominguez D R C, Korutcheva E, Theumann W K and
Erichsen Jr. R, Flow diagrams of the quadratic neural network,
     {\it Lecture Notes in Computer Science}, {\bf 2415}, (2002) 129.
\bibitem{BM00} Boll\'e D and Massolo G, 
       Thresholds in layered neural networks with variable activity, 
       {\it J. Phys. A: Math. Gen.} {\bf 33} (2000) 2597.
\bibitem{HKP91} Hertz J, Krogh A and Palmer R G, 
       {\it Introduction to the Theory of Neural Computation}, 
         Addison-Wesley,  Redwood City (1991).
\bibitem{DKM} Domany E, Kinzel W and Meir R,
         Layered Neural Networks,  
       {\it J.Phys. A: Math. Gen.} {\bf 22} (1989) 2081.
\bibitem{B04} Boll\'e D, 
Multi-state neural networks based upon spin-glasses: a biased overview, in {\it Advances in Condensed Matter and Statistical Mechanics} eds. Korutcheva E and Cuerno R., Nova Science
Publishers, New-York,(2004 p. 321-349. 
\bibitem {NBP98} Nadal J-P, Brunel N and Parga N,
     Nonlinear feedforward networks with stochastic outputs: infomax implies redundancy
     reduction, 
    {\it Network: Computation in Neural Systems} {\bf 9} (1998) 207.
\bibitem{ST98} Schultz S and Treves A,
     Stability of the replica-symmetric solution for the information conveyed by a neural
    network. 
    {\it Phys. Rev. E } {\bf 57} (1998) 3302.
\bibitem{B90} Blahut R E,  
     {\it Principles and Practice of 
     Information Theory}, Reading, MA: Addison-Wesley (1990).
\bibitem{S48}  Shannon C E,  
    A mathematical theory for  communication, 
    {\it  Bell Systems Technical Journal} {\bf 27} (1948) 379.
\bibitem{SSP96} Schwenker F, Sommer F T and Palm G, 
     Iterative retrieval of sparsely coded associative memory patterns, 
    {\it Neural Networks} {\bf 9} (1996) 445.
\bibitem{A77} Amari S, 
   Neural theory and association of concept information, 
   {\it Biol. Cyber.} {\bf 26} (1977) 175.
\bibitem{AM88} Amari S and Maginu K, 
    Statistical neurodynamics of associative memory,
    {\it Neural Networks} {\bf 1} (1988) 63.
    
\end{thebibliography}
\end{document}